# Flow Past Stationary and Oscillating Airfoil at Low Reynolds number Using Sharp Interface Immersed Boundary Approach

*Pradeep Kumar S, Ashoke De\**

*Department of Aerospace Engineering, Indian Institute of Technology Kanpur, 208016, Kanpur, India*

\*Address all Correspondence to Ashoke De
E-mail: ashoke@iitk.ac.in

*The present study reports on flow past airfoils (stationary and moving) using sharp interface immersed-boundary approach. Non-boundary conforming approach like immersed-boundary method offers a viable alternative over traditional boundary conforming approach by allowing us to model flow past arbitrarily complex shapes, by eliminating the need to re-grid the flow domain as the body exhibits motion. We present flow past a NACA 0012 airfoil at stationary conditions as well as exhibiting pitching motion. Evolution of vortex dynamics and wake structures are presented to show that the developed sharp interface immersed-boundary approach captures the flow physics of dynamic stall accurately. Moving body problems involving immersed-boundary approach usually encounter the issues of spurious oscillations and mass conservation. This is handled through a field extension strategy based on ghost cell approach, which allows for extrapolating the flow field value onto the ghost nodes, ensuring smooth temporal transition as the immersed surface moves through time. The results presented here show excellent agreement with the experimental results found in the literature.*

## 1. Introduction

The interests in biomimetic propulsive systems have grown tremendously in the past decade because of its relevance to several military and civilian applications. Of particular interest is the aerodynamics of flapping foil operating at low speeds. An important flow phenomenon associated with flapping wings is 'dynamic stall' (Akbari and Price 2003). As the airfoil pitches through its static stall angle, the normal force continues to increase beyond its maximum value corresponding to the unstalled value. This increase in normal force is attributed to the formation of leading-edge vortices (LEV). LEVs travel along the airfoil surface as it grows until it reaches maximum pitch amplitude. During the stroke reversal, the growth stops, and LEVs are shed. This complex flow phenomenon is further complicated by the fluid-structure interaction that takes place between the vortical structure and wing body. Capturing these complex phenomena calls for simple, efficient and robust computational models that can provide a realistic assessment and shed new insights into the flow physics. As an initial step towards attaining this objective, the idea of a sharp interface immersed-boundary approach is utilized.

Interests in non-boundary conforming approaches have increased in recent times as it has provided a viable alternative to boundary confirming methods by enabling the realistic representation of complex geometries. The procedure also eliminates the requirement for re-gridding the entire mesh domain where the object moves with time. The sharp interface (Fadlun et al. 2000, Kim et al. 2001, Udaykumar et al. 2001, Gilmanov and Sotiropoulos 2005, Yang and Balaras 2006, Mittal et al. 2008, Seo and Mittal 2011, Kumar and Roy 2016) unlike diffused interface approach (Peskin 2002, Kumar et al. 2015) allows for accurate representation of immersed geometry by allowing exact imposition of boundary conditions providing a better resolution of the boundary layer near the immersed surface.

One major issue faced by the sharp interface immersed-boundary approach when encountering the moving body problems is the spurious temporal oscillations of pressure and force. To mitigate this problem, we have implemented a field extension strategy based on ghost cell approach. For validation, we have simulated a case involving inline oscillating cylinder at Re=100 in a medium, where the flow is initially at rest, and compared the predictions with the measurements. Later, for better insight about the model that we have developed, a systematic study is carried out involving flow past NACA0012 airfoil. Initially, we present the flow past a stationary airfoil (NACA0012) at Re = 5300 over a range of angles of attack ($\alpha = 10°$ to $90°$). Then, we report pitching airfoil (NACA0012) undergoing a motion at three different pitching locations for a given reduced frequency f*= 0.5 at Re = 3000.

## 2. Numerical Methods

## 2.1 Flow Solver

The unsteady, 3D Favre averaged Navier Stokes equations are solved using density based Finite Volume solver. The multi-block solver uses co-located grid structure, and equations are solved in generalized curvilinear coordinates. For simulating low Mach number and incompressible flows, the preconditioning technique is used. The convective fluxes are discretized using Low Diffusion Flux Splitting Scheme (LDFSS). The viscous fluxes are discretized using a central difference scheme. A dual time stepping approach is used for time marching. The physical time steps are discretized using second order backward three-point differencing while local pseudo time is discretized using Explicit Euler differencing. Parallel processing uses MPI. Further details about the solvers can be found in the literature (Seshadri and De 2017, De and Acharya 2012, Das and De 2015, Das and De 2016, Arya and De 2019 ).

## 2.2 Sharp Interface Immersed-Boundary Approach

In this approach, the immersed surface is represented by an unstructured triangular mesh. All the cell centers that lie within and outside those surfaces are classified by calculating sign of the scalar product between surface normal of the triangular elements and the position vector of nearby Cartesian grid points. The nodes are then tagged as fluid, solid and immersed boundary (IB) nodes. IB nodes are nearest neighbor cell center nodes from the immersed surface. These nodes act as forcing points whose values $\varphi$ (=velocity and pressure) are reconstructed according to the boundary conditions provided by the surface. The reconstructed values are computed using quadratic interpolation stencil ( $\varphi = C_1 n^2 + C_2 n + C_3$ ) along the local normal (*n*) drawn from the IB node (see BI-B line in **Fig. 1**). The three unknown co-efficients in the quadratic stencil ( $C_1, C_2$ and $C_3$ ) are computed by solving three equations formed using the information from two fluid points and one body intercept (BI) point (Seshadri and De 2017). At the body intercept (BI) Dirichlet boundary condition is prescribed for velocity which ensures no-slip condition. Similarly for pressure, Neumann boundary condition is utilized. In case of stationary object, zero normal pressure gradient ensures non penetration condition. In case of moving bodies the normal pressure gradient is set equal to the material acceleration of immersed surface nodes (this expression results from applying momentum equation along the normal). The fluid points (for example A,B,F in **Fig.1**) need not always coincide with cell center nodes. Flow variables at such points are calculated using Inverse Distance Weighted interpolation (IDW). The approach is similar to the works of Gilmanov and Sotiropoulos (2005); Kumar and Roy (2016).

Moving body problems in sharp interface immersed-boundary approach faces issues of mass conservation and spurious oscillation because of the violation of geometric conservation of mass (Kumar and Roy 2016). The sudden and abrupt role change of grid nodes, from solid to fluid, fluid to solid leads to unphysical values entering spatial and temporal terms. This accumulated error appears as spurious, high-frequency oscillations. To suppress these oscillations, a field extension strategy based on Ghost Cell methodology introduced by Yang and Balaras (2006) is used. According to this, the ghost node G immediately beneath the immersed surface (see G-F line **Fig 1**) is reconstructed based on the flow field outside such that it satisfies the immersed surface boundary conditions. The extrapolation is done using similar quadratic stencil discussed earlier involving two fluid points and one body intercept point (BI) along the normal drawn from the ghost node. Such ghost nodes when it emerges as IB nodes or fluid nodes as the body moves provide a smoother transition.

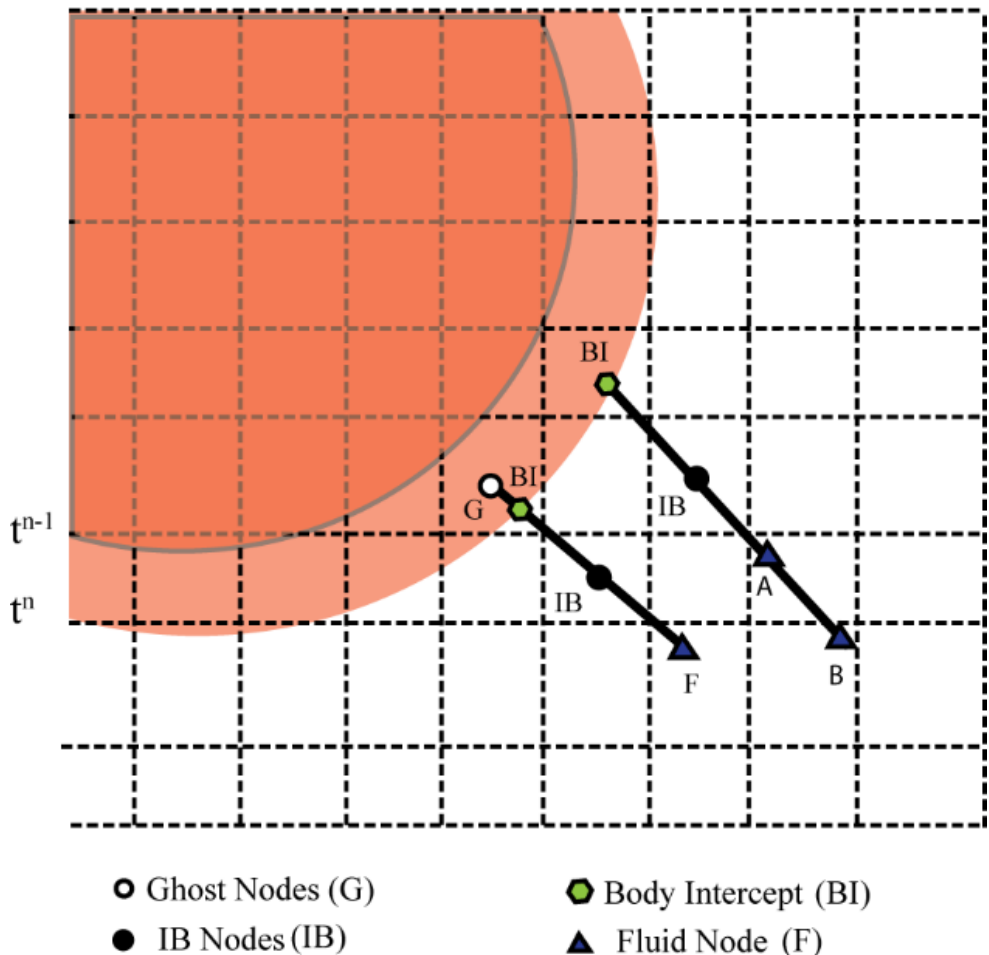

**Figure 1 :** Schematic of sharp interface immersed-boundary approach and field extension Approach.

## 2.3 Force Calculations

The forces on the body are calculated by integrating the pressure $P$ and viscous stresses $\tau$ over the immersed surface $\Sigma$ using the expression

$$F_i = \int_\Sigma \left[-P\delta_{ij} + \tau_{ij}\right] n_j d\Sigma \tag{1}$$

Where $F_i$ denotes force in $i^{th}$ cartesian co-ordinate. Here also, the similar quadratic interpolation stencil is invoked to obtain the pressure and viscous stresses on the immersed surface.

## 3. Results and Discussions

### 3.1 Flow Past Stationary Airfoil

Here we report the predictions of flow past a 2D NACA 0012 airfoil for a Reynolds number of 5300 at different angles of attack, i.e. $\alpha = 5°, 10°, 20°, 40°, 60°, 80°, 90°$. The Re is based on airfoil's chord length ($c$). The Cartesian fluid domain is of the size *40c x 30c*. A uniform grid of size $c/\Delta x = 84$ is adopted locally while grids are stretched non uniformly away from the airfoil.

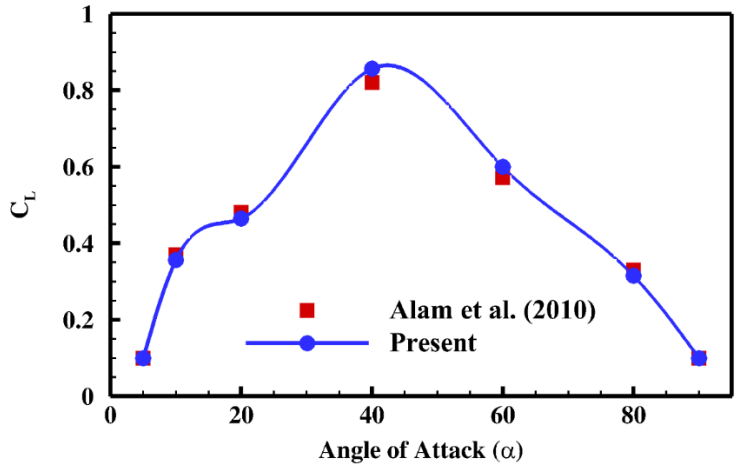

**Figure 2** : mean $C_L$ Vs. $\alpha$ curve for NACA0012 at Re = 5300 compared with Alam et al. (2010)

**Fig.2** shows the mean lift coefficient variation at different angles of attack in comparison with the experimental work of Alam et al. (2010). Alam et al. (2010) observed that at this particular Reynolds number (Re = 5300), the stall is absent up to $\alpha = 45°$, leading to a monotonic increase in the lift co-efficient. The present simulation too has captured this exciting behavior. The absence of stall is attributed to the laminar nature of the separation bubble even at the higher angles of attack at this Re.

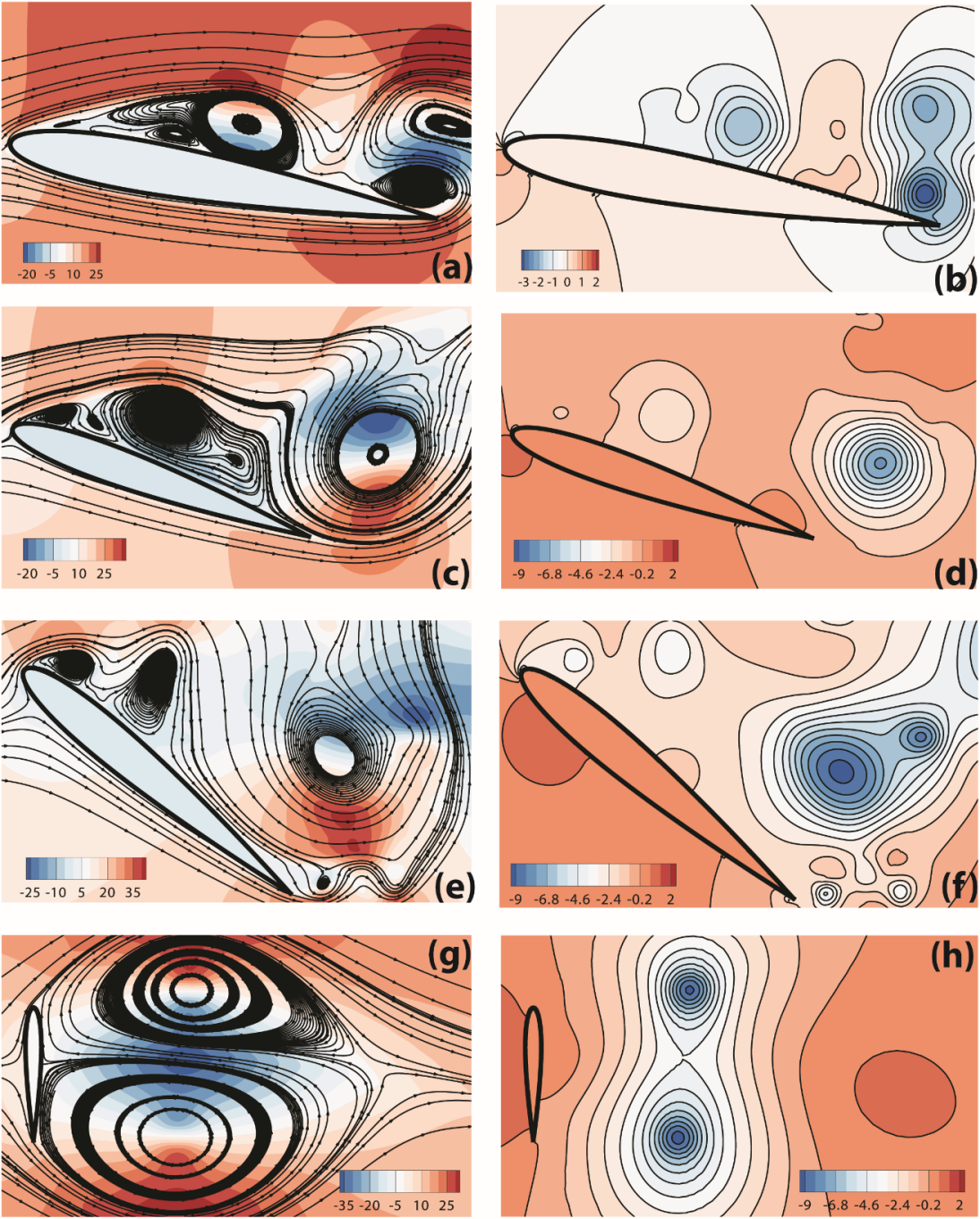

**Figure 3** : Flow past stationary NACA0012 airfoil (Re = 5300) at increasing angles of attack ( $\alpha = 10°, 20°, 40°, 90°$ ). First column (a,c,e,g) shows streamline plot. Second column (b,d,f,h) shows pressure contour plot. The plots correspond to flow through time t*=50

**Figure 4:** Contours of normalized vorticity $\omega^* = \omega c / U_\infty$ at t* =50 : (a) $\alpha = 10°$ (b) $\alpha = 20°$ (c) $\alpha = 40°$ (d) $\alpha = 90°$

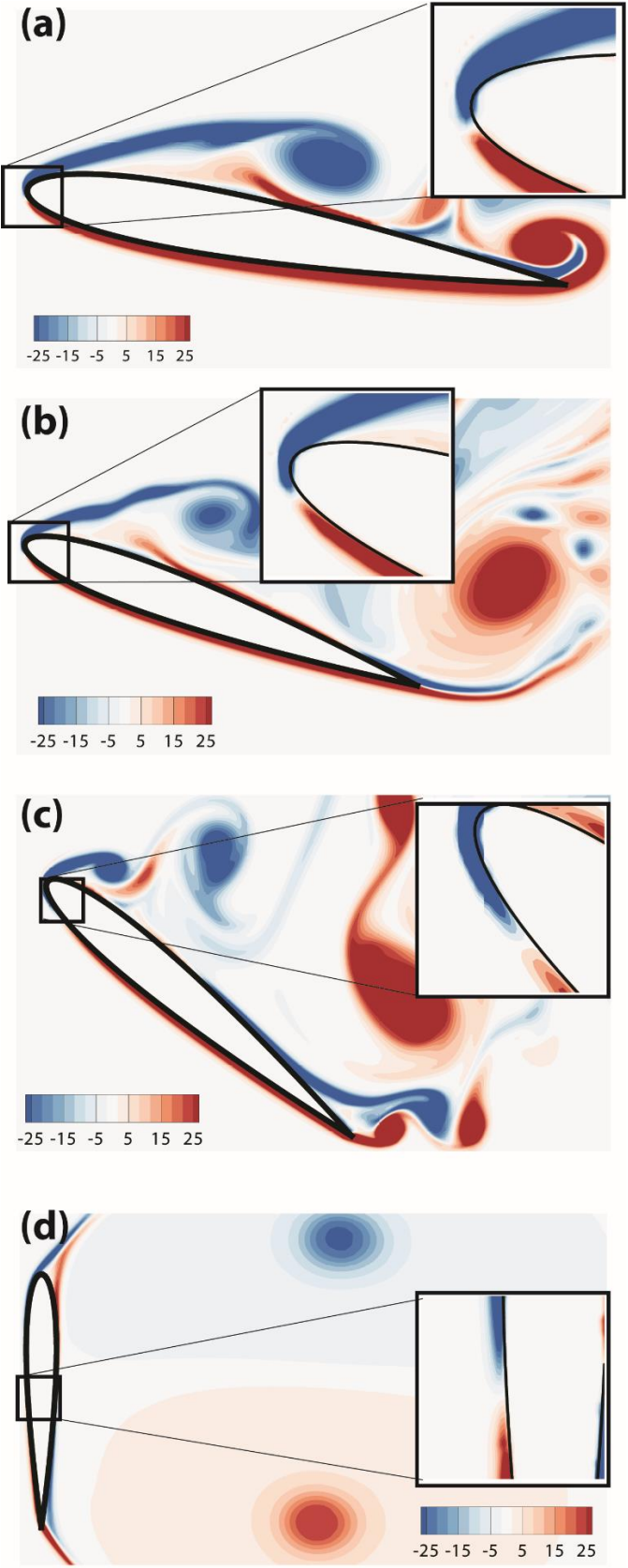

Resolving the boundary layer phenomenon has always been a challenge for the immersed-boundary approach. If the sharp interface is not maintained, the mass conservation will be violated. This will, in turn, lead to unphysical velocity and pressure values near the immersed surface affecting the accuracy as well as flow physics of the problem. The vorticity distribution calculated from the unphysical velocity will inevitably affect the phenomena like flow separation.

Fig.3 presents instantaneous streamline and pressure contour plot corresponding to the angles of attack $\alpha = 10°, 20°, 40°, 90°$. These plots correspond to flow through time $t^* = tU_\infty/c = 50$. The plots show a sharp representation of airfoil at its surface and smooth variation flow variables near the trailing edge. Maintaining sharp representation near the trailing edge is difficult. These results show that our quadratic reconstruction schemes help in retaining sharp features near the trailing edge. Notably, the flow corresponding to $\alpha = 40°$ (Fig.3e) shows that the separation bubble is laminar and flow reattaches itself near the mid-chord. This is consistent with experimental observation of Alam et al. (2010).

Though the airfoil is stationary, the flow is unsteady. Fig.4 depicts the normalized vorticity plot corresponding to the instant t*=50. Enlarged view of forward stagnation point (region where $\omega^* = 0$) is shown for the angle of attacks presented. At $\alpha = 0°$, forward stagnation point is expected to be right at the nose. As the angle of attack increases, the forward stagnation point shifts from its leading edge nose point toward the trailing edge and ends up near the middle of two edges for $\alpha = 90°$. This is consistent with the Alam et al. (2010) observation. Also note that the trailing edge vorticity is smooth and clean, reinforcing our earlier observation that sharp features are maintained in the trailing edge.

## 3.2 Inline Oscillating Cylinder in a Quiescent Medium

As discussed in section 2.2, we have utilized a field extension strategy based on ghost cell approach for handling moving body problems. For validation, we have simulated an inline oscillating cylinder in a quiescent medium. The case is simulated for $Re = UD/\nu = 100$ and Keulegan-Carpenter number $KC = 2\pi A_e/D = 5$ where *D* is the diameter of the cylinder, $\nu$ is the kinematic viscosity, *U* is the maximum velocity with which cylinder oscillates, and $A_e$ is the amplitude of cylinder oscillation. These parameters correspond to (Dütsch, Durst et al. 1998, Guilmineau and Queutey 2002). The domain is $50D \times 30D$ where D is the diameter of the cylinder.

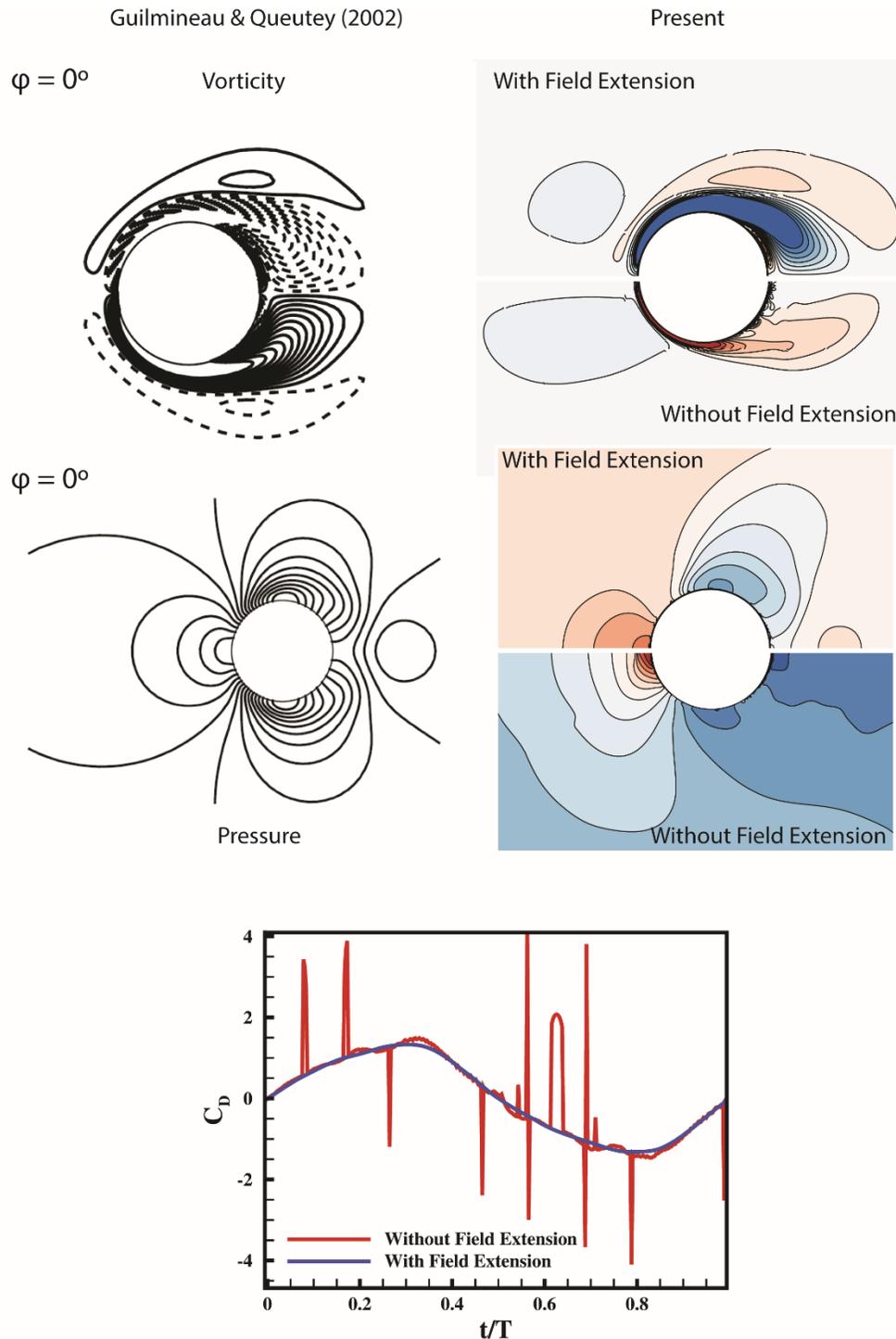

**Figure 5:** Comparison of the flow field of inline oscillating cylinder corresponding to phase $\varphi = 0°$. The first column shows the vorticity and pressure distribution obtained by Guilmineau and Queutey (2002) through simulation. The second column shows flow field results corresponding to present simulation with and without field extension approach. The time history of drag co-efficient for one cycle and compared with without field extension approach.

The expression gives the sinusoidal oscillation along x-direction as:

$$X(t) = -A_e \left[\sin(\varphi)\right] \quad (2)$$

Here phase $\varphi$ is given by

$$\varphi = 2\pi f_e t \quad (3)$$

where $f_e$ is the frequency of the oscillating cylinder. The upper and lower boundaries are set to no-slip wall boundary conditions while symmetry boundary condition is applied for the pressure and velocity at the inlet and outlet boundaries.

Fig.5 exhibits flow field results corresponding to phase $\varphi = 0°$. The flow field results obtained with field extension and without field extension strategy are compared with the flow field results of Guilmineau and Queutey (2002). The results corresponding to without field extension (in Fig.5) shows a chaotic boundary layer vorticity and spurious pressure contour. When compared to Guilmineau and Queutey (2002), the results shown are entirely erroneous. On the other hand, field extension strategy resolves the boundary layer, captures the pressure distribution accurately. The time history of drag co-efficient comparing with and without field extension is also presented in Fig.5. The results show that field extension strategy suppresses spurious oscillations.

Fig.6 presents the vorticity and pressure contours around the cylinder surface. The vorticity contour exhibits the existence of two stagnation point throughout the motion. One at the front of the cylinder and another at the rear end. The flow is characterized by stable and symmetric vortices that shed periodically. As the oscillating cylinder moves forward ($\varphi = 0°$), an upper and lower boundary layer develops in front of the cylinder. This boundary layer separates at the same upper and lower position at the cylinder surface. The separated flow results in two counter-rotating vortices of the same structure and strength ($\varphi = 95°$). As the oscillating cylinder reaches its maximum forward position, the vorticity creation stops and flow reversal begins as the cylinder begins its backward motion. As the cylinder moves backward, an upper and lower boundary layer develops ($\varphi = 193°$) as observed in forwarding motion. The separated flow forms two counter-rotating vortices as the cylinder moves further back ($\varphi = 288°$). The pressure distribution corresponding to these phases are also presented alongside in Fig.6. The results are in good agreement with Dütsch et al. (1998); Guilmineau and Queutey (2002).

**Figure 6 :** Vorticity and Pressure contour around the cylinder surface over different phases of oscillation

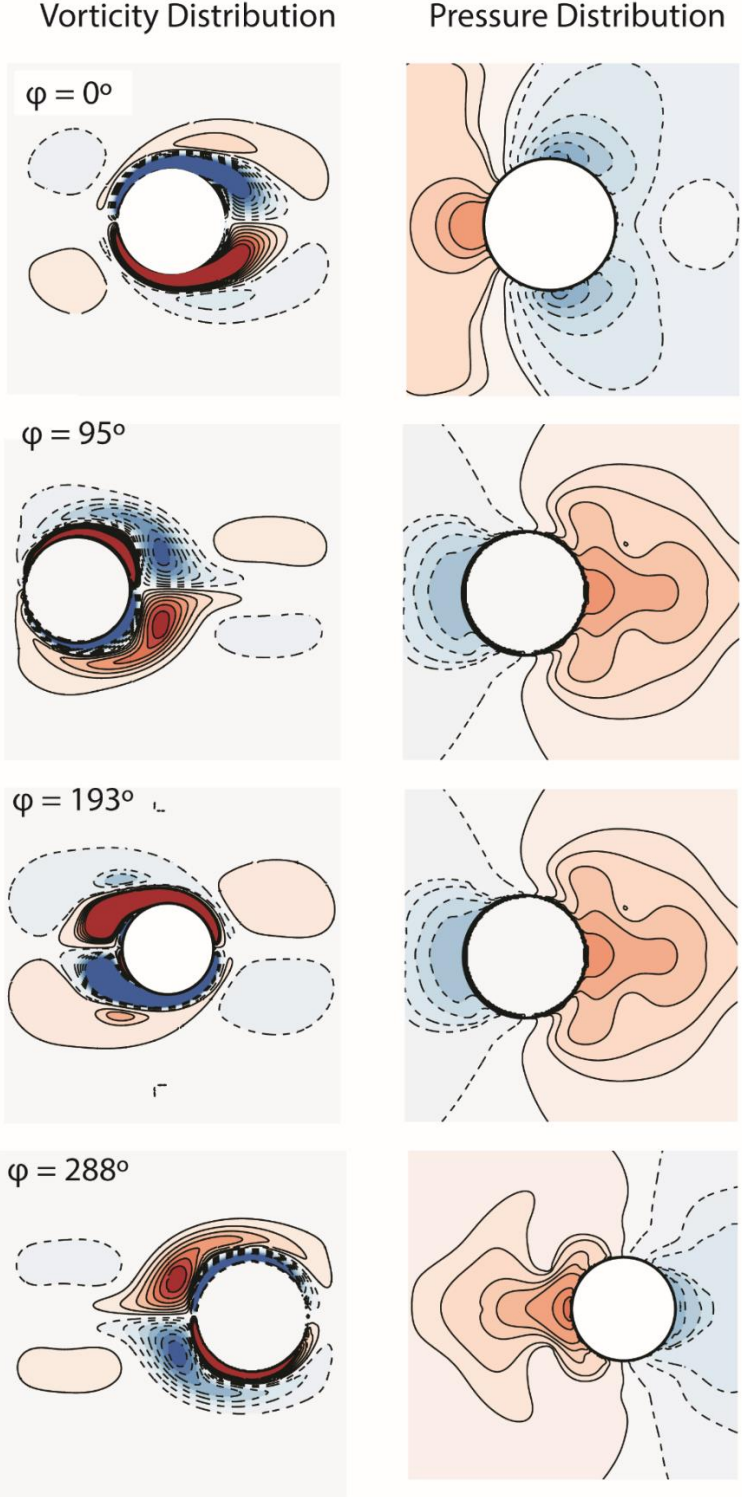

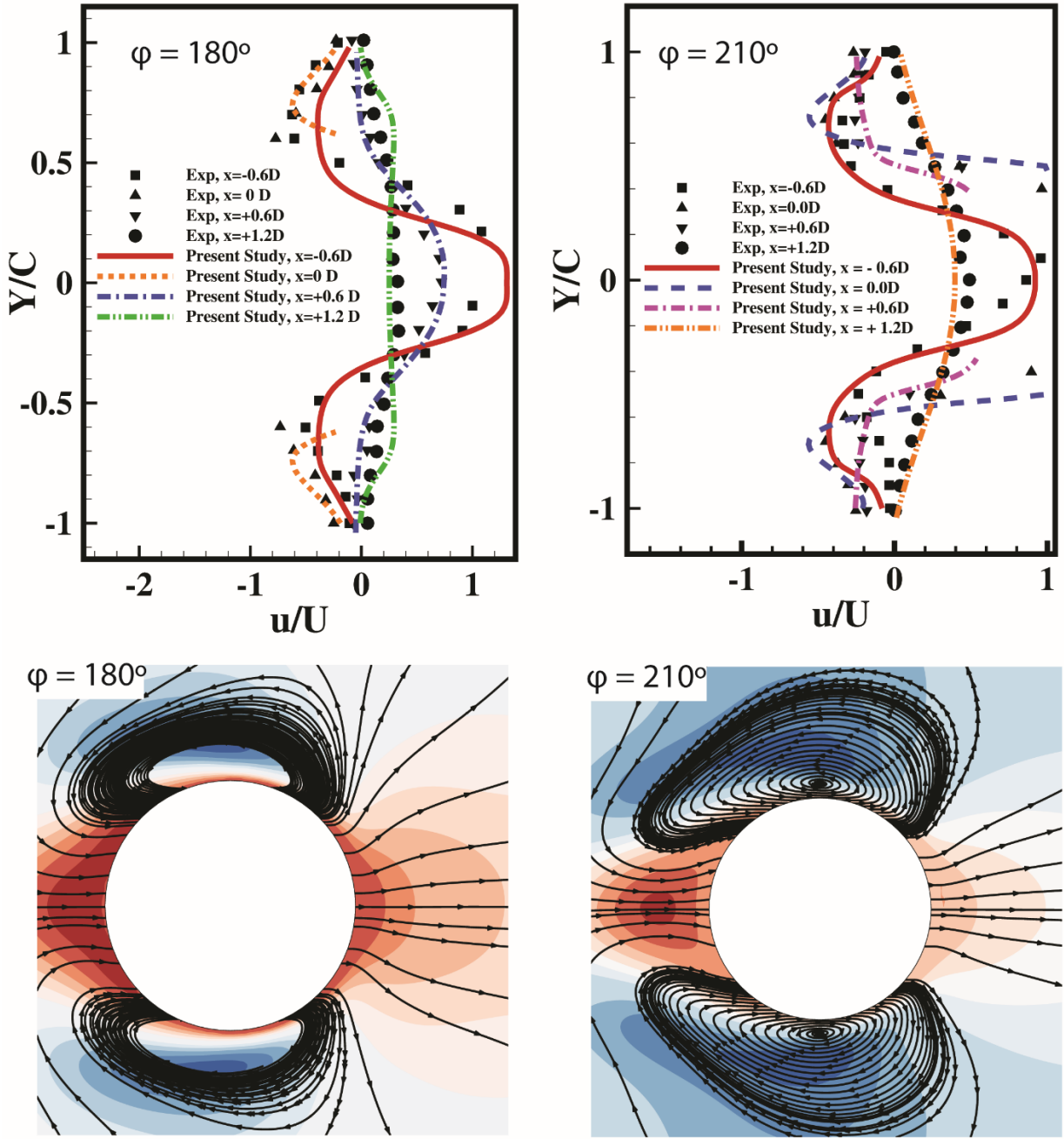

**Figure 7 :** First row shows Velocity profile corresponding to phase $\varphi = 180°\ \&\ 210°$ compared with the experimental results of (Dütsch, Durst et al. 1998). The streamline plot for the two phases are shown in the second row

Further, the velocity profile comparison (Fig. 7) with the experimental results obtained by Dütsch et al. (1998) for two phases ($\varphi = 180°$ & $210°$) are also found to be in good agreement. The corresponding streamline plots are also depicted in the same graph.

## 3.3 Dynamic stall of Pitching NACA0012 airfoil

This section presents a 2D simulation of NACA0012 airfoil exhibiting pitching motion. The parameters are chosen from the PIV study of Ohmi et al. (1991). The Reynolds number (based on chord length) is 3000. The mean incidence of the airfoil is $\bar{\alpha} = 30^o$, and the angular amplitude $\Delta\alpha$ is 15°. The expression gives the instantaneous angle of attack governing the pitching motion is $\alpha = \bar{\alpha} - \Delta\alpha \cos(2\pi f t)$. The Eulerian fluid domain is of size 40cX30c with 425x317 nodes in the X-Y Plane. The reduced frequency is given by $f^* = fc/2U_\infty$. The grid is uniformly refined locally near the immersed boundary. The objective of this section is to study the effect of pitch axis location on vortex dynamics. NACA0012 airfoil pitching about three locations (one-third, half and two-thirds of the chord, i.e. x/c = 1/3, 1/2, 2/3) at reduced frequency 0.5 are considered in this section.

Figs.8-9 report normalized instantaneous vorticity ($\omega^* = \omega c / U_\infty$) contour plots of airfoil pitching about one-third of the chord (x/c = 1/3). The figures depict a detailed evolution of vortices over two cycles of the pitching motion. The airfoil begins its pitching motion from minimum incidence 15° at t =0. Shortly after the beginning of the motion, it is noticed (Fig.8) that a vortex forms at the trailing edge (t = 0.1T). As the pitching motion progresses, the trailing edge vortex rolls up and sheds into wake region(t = 0.3T). When the airfoil reaches the end of its upstroke (t =0.5T), a local thickening of the boundary layer at the leading edge is observed. As the down-stroke begins, the thickened boundary layer at the leading edge rolls up to form leading edge vortex (t = 0.7T). This leading-edge vortex is shed as the airfoil reaches the end of the down-stroke (t =1.0T). Also, notice that trailing edge forms a clockwise vortex (t=0.8T) that is shed into the wake as it completes the cycle. Fig.9 shows that leading edge vortex shed from the first cycle interacts with the boundary layer of opposite vorticity on the surface forming a dipole. The alternately shed trailing edge vortices also form dipole in the wake. The dipole formed at the surface is shed when the airfoil reaches the end of the upstroke of the second cycle (t=1.5T). The shed dipole is convected downstream. The vortex dipole interacts with the trailing edge boundary layer and collapses onto it when the airfoil reaches the end of its down-stroke (t=2.0T). This phenomenon is called as wake capture and is observed in foils oscillating in high frequencies. Birds and insects exploit this to extract energy to improve its flight performance. The end of the second cycle is also marked with the formation of another leading edge vortex.

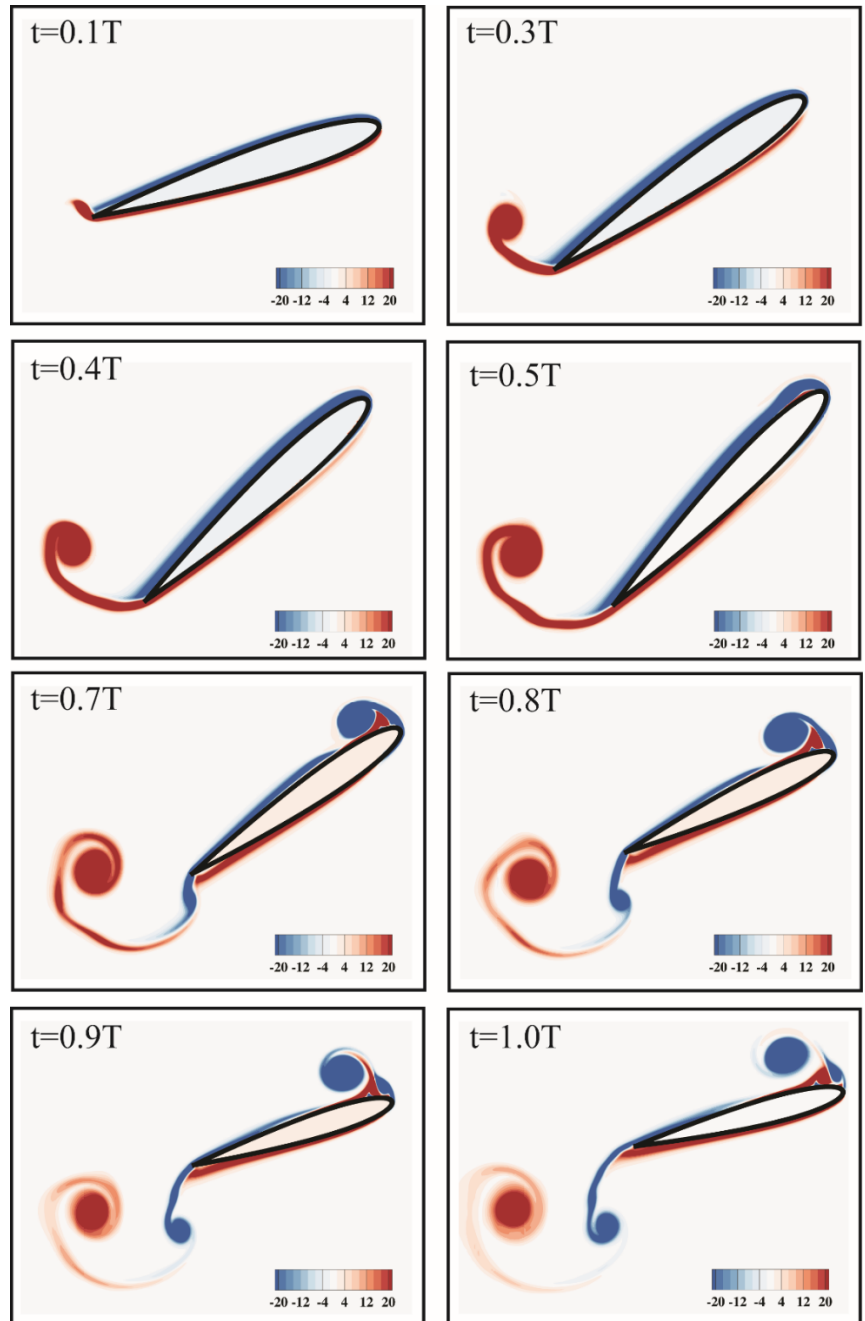

**Figure 8 :** Normalized vorticity contours plots of NACA0012 pitching about one-third of the chord. The plots correspond to the first cycle. 'T' refers to the time period of the pitching motion.

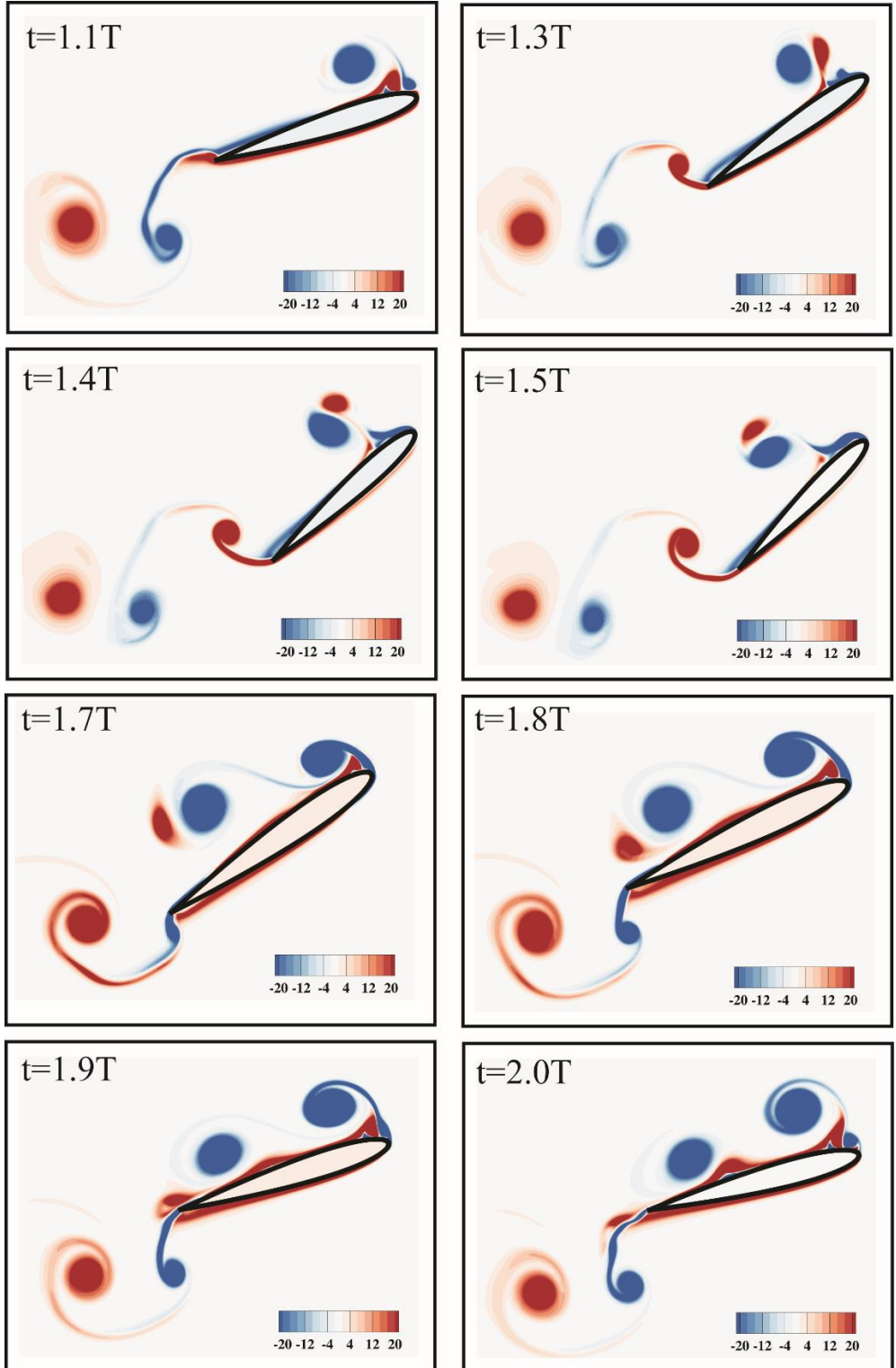

**Figure 9:** Normalized vorticity contours plots of NACA0012 pitching about one-third of the chord. The plots correspond to the second cycle. 'T' refers to the time period of the pitching motion.

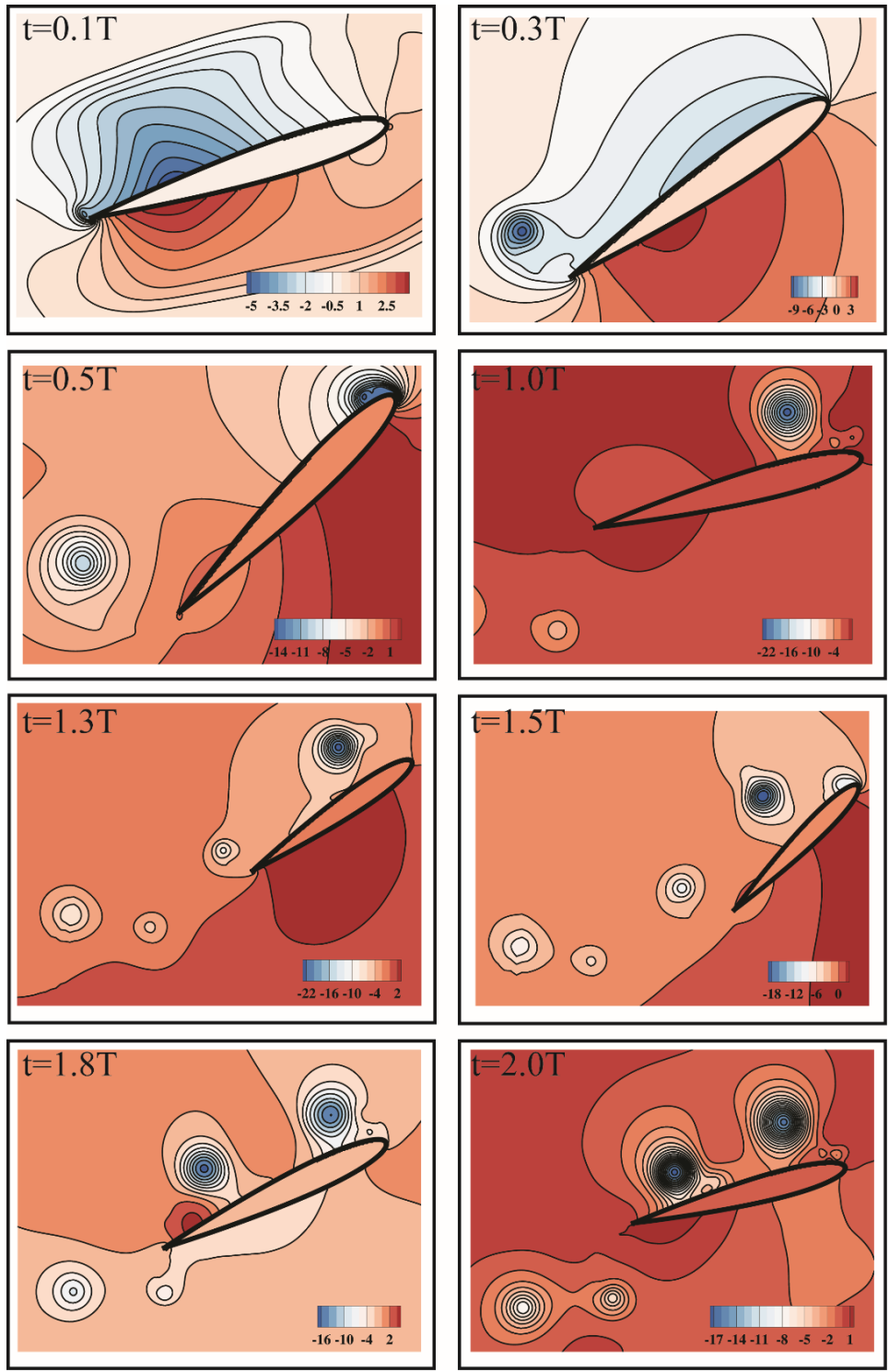

**Figure 10:** Pressure coefficient contour plots of NACA0012 pitching about one-third of the chord.

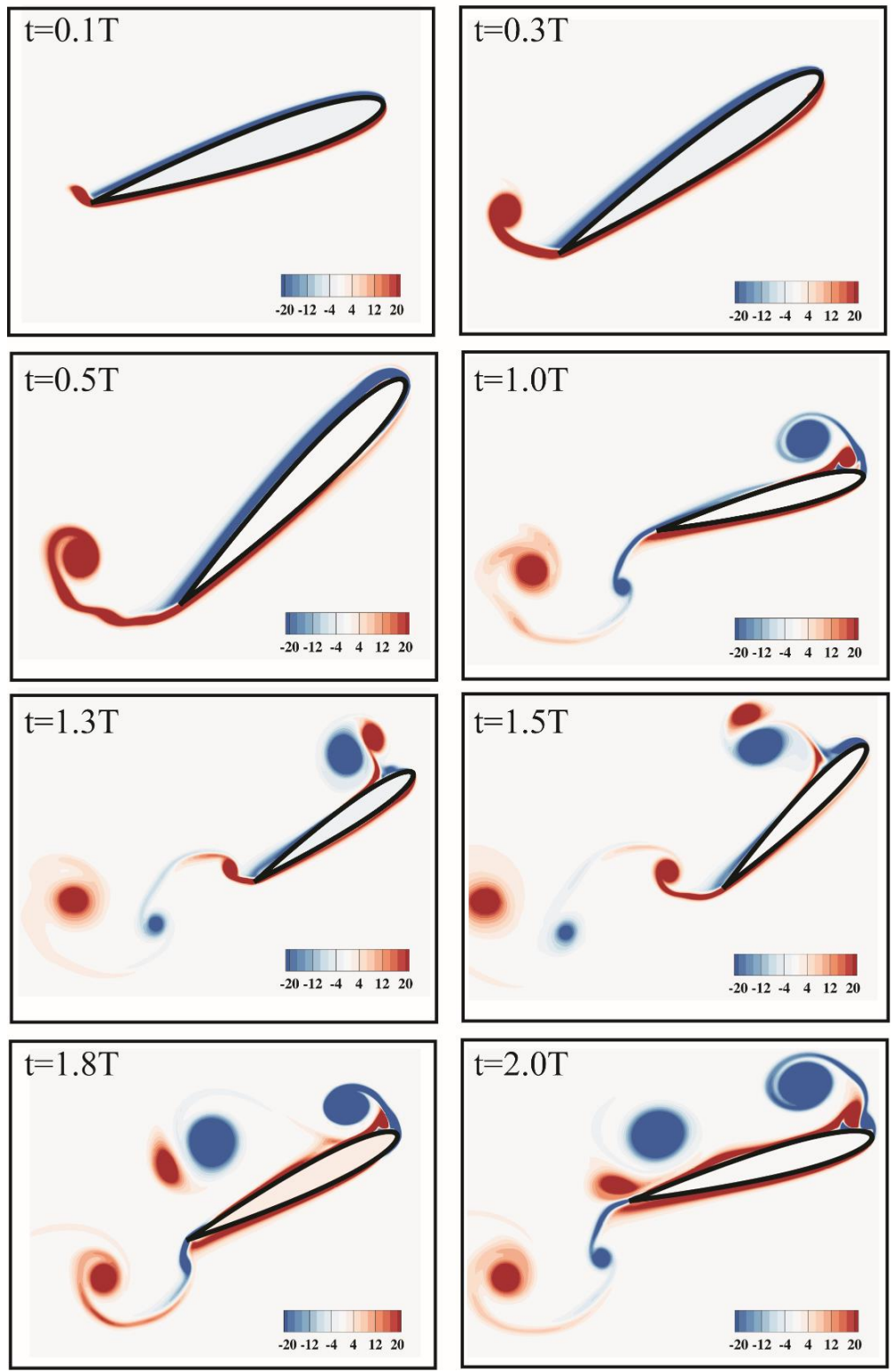

**Figure 11:** Normalized vorticity contours plots of NACA0012 pitching about the half chord.

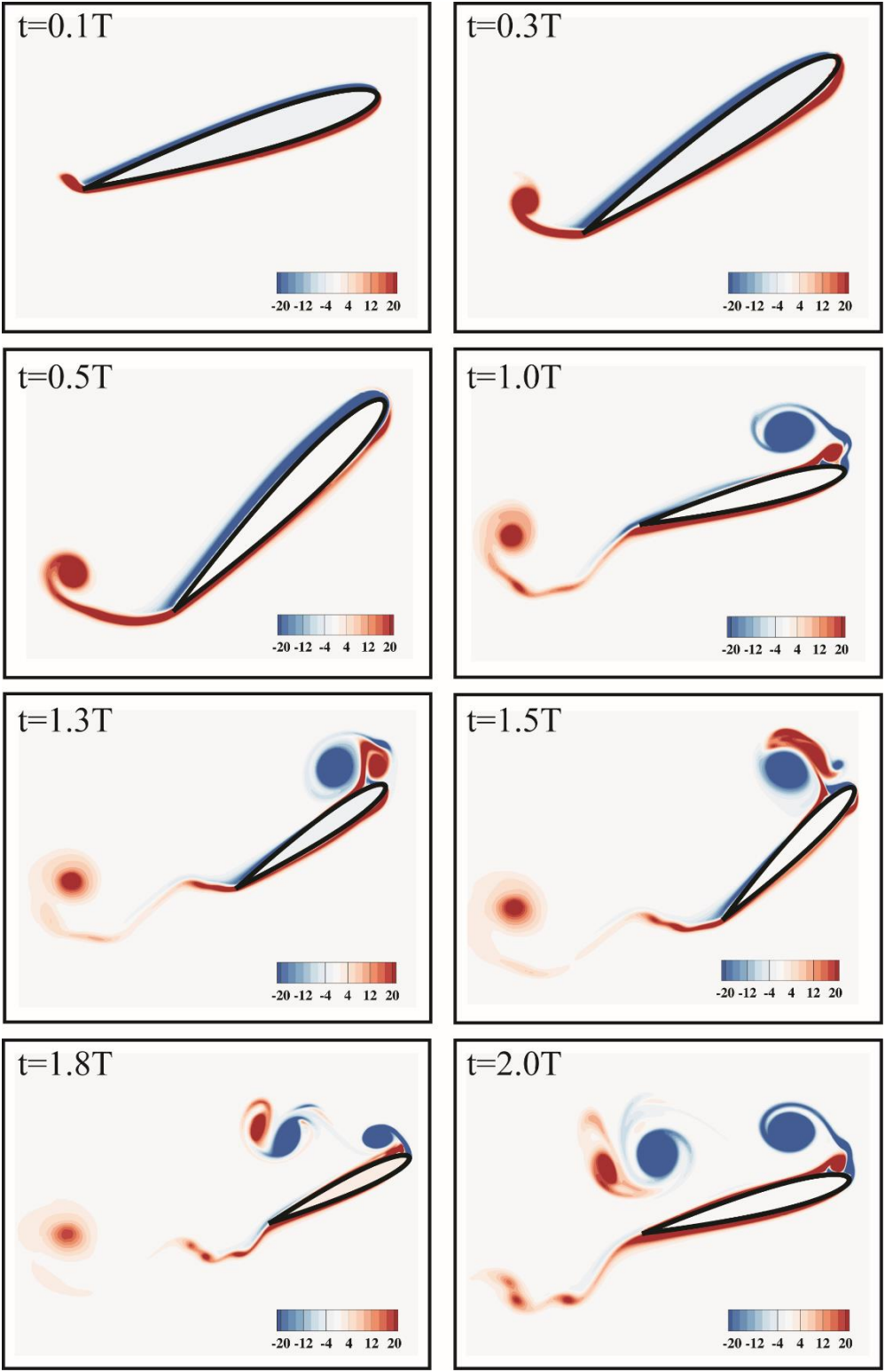

**Figure 12:** Normalized vorticity contours plots of NACA0012 pitching about two third of the chord.

Ohmi et al. (1991) too observed that trailing edge generates clockwise and counter clockwise vortices alternately at this reduced frequency. Another observation that growth of leading-edge vortex is aided by the downstroke motion (t=1.0T, 2.0T), while suppressed by the upstroke motion (t = 0.5T, 1.5T) is also well captured by the present simulation. Fig.10 shows pressure co-efficient contour plot over two cycles. It is to be noted that the pressure distribution is smooth near the airfoil surface and sharp representation of the surface is maintained at all the time instances.

Fig.11 reports normalized instantaneous vorticity contour of NACA0012 airfoil pitching about its half chord location. The vortex dynamics is very similar to that of the case (Fig.8-9) where the pitch axis is located at one-third of the chord. Generation of clockwise and counter clockwise trailing edge vortices, growth and suppression of leading-edge vortex concerning the motion of trailing edge are observed in this case too.

Fig.12 corresponds to the case where the pitching axis is located about two third of the chord. The vortex dynamics near the leading edge is very similar to that of the earlier examples. Note that the dipole formed near leading edge convects downstream like in the previous cases but do not interact with the surface at the trailing edge even at the end of the second cycle(t=2.0T). At the trailing edge, the alternate shedding of vortices is absent. Except for generation of starting vortex which is shed at t = 0.5T. For the rest of the one and a half cycle, no trailing edge vortex is formed.

Figs.13-16 depict a comparison of present simulation with that of experimental results of Ohmi et al. (1991) at four-time instances (t =0.5T,1.0T,1.5T,2.0T). At t=0.5T, results from Ohmi et al. (1991) shows the presence of small leading-edge vortex and a starting vortex shed from the trailing edge (Fig.13a). The plot corresponds to the case whose pitch axis is located at one-third of the chord. When the pitch axis location is moved down to half chord, the leading edge vortex is not observed (Fig.13b). The strength of the shed vortex is also less compared to one-third of chord case. Fig.13c-d shows good agreement with the experiments. When the pitch axis location is further moved down to two third of the chord (Fig.13e), a similar trend is observed. The strength of the shed vortex is much less compared to two earlier cases. No leading edge vortex is present.

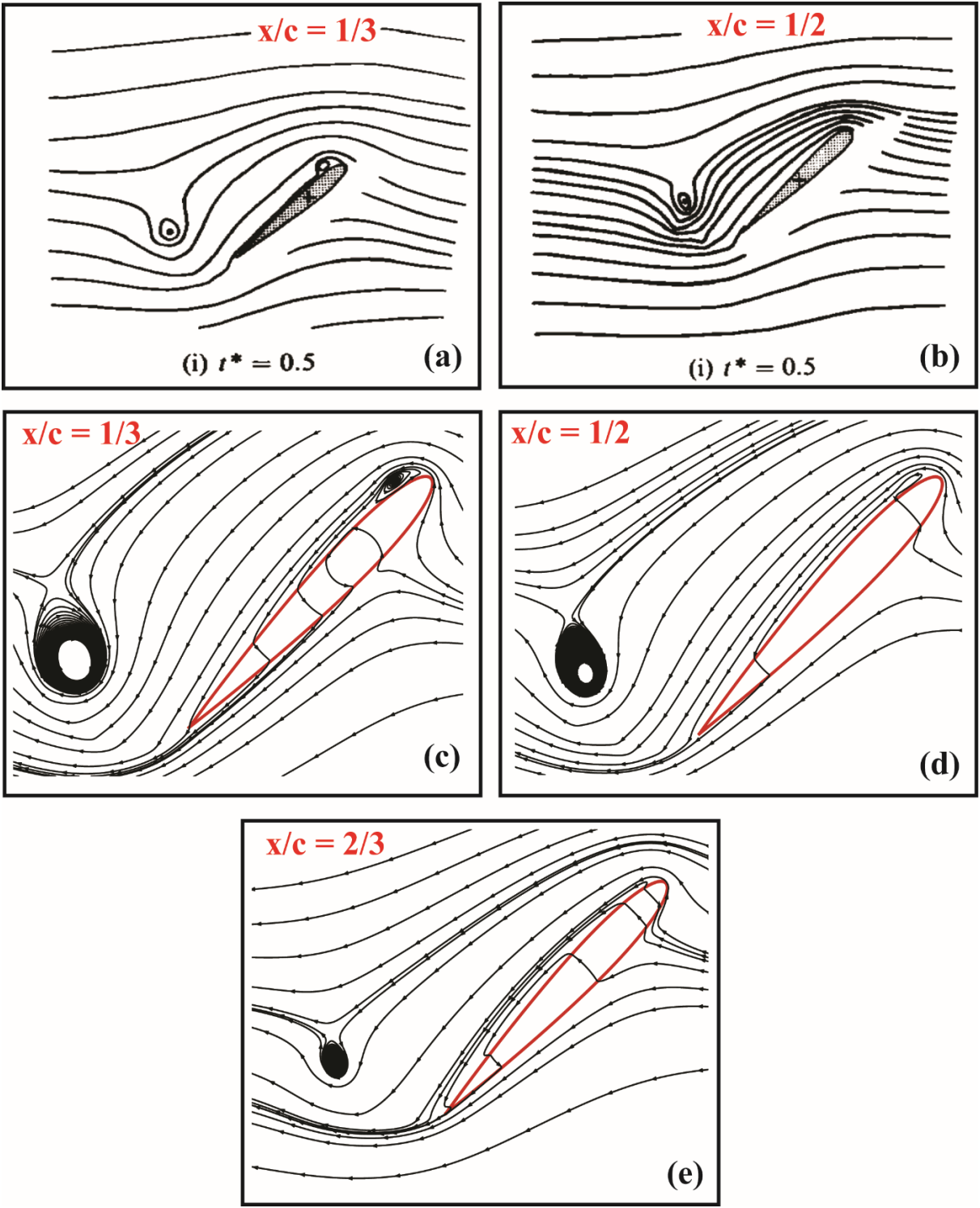

**Figure 13:** Streamline pattern for time instance t=0.5T. (a,b) Ohmi et al's experimental observation [11]. (c,d,e) Results corresponding to three different pitching location (x/c)

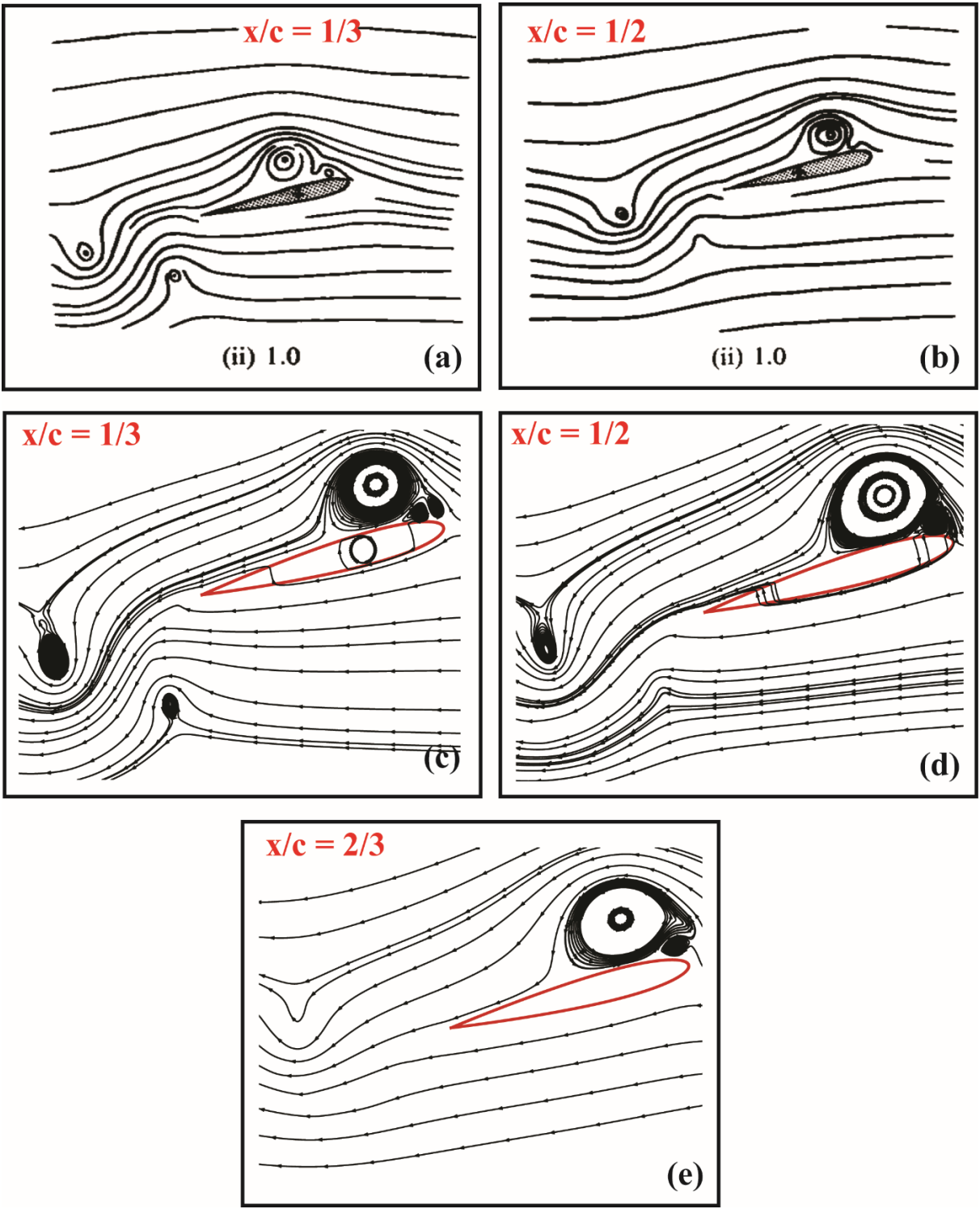

**Figure 14:** Streamline pattern for time instance t=1.0T. (a,b) experimental observation by Ohmi et al (1991). (c,d,e) Results corresponding to three different pitching location (x/c)

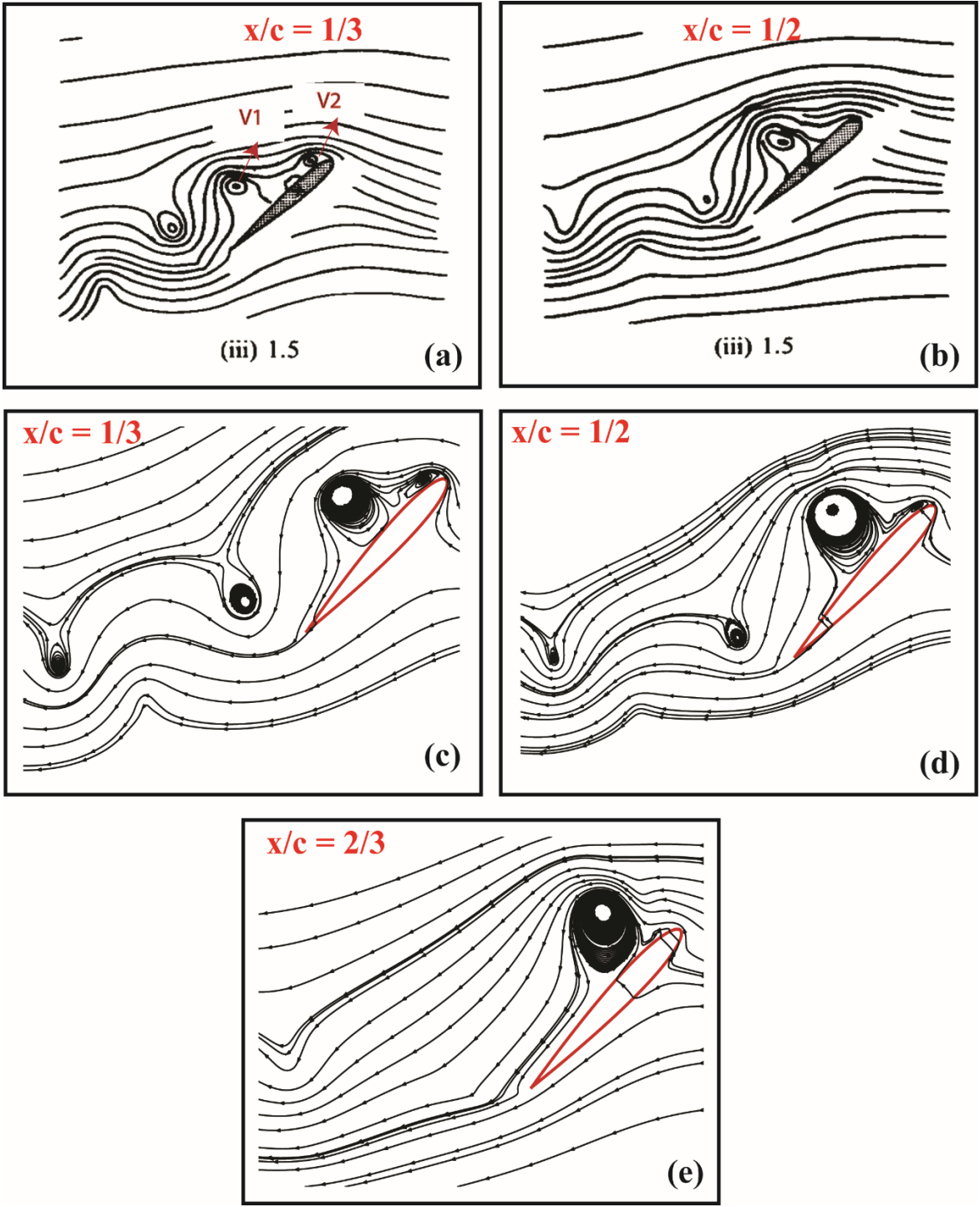

**Figure 15:** Streamline pattern for time instance t=1.5T. (a,b) experimental observation Ohmi et al. (1991). (c,d,e) Results corresponding to three different pitching location (x/c)

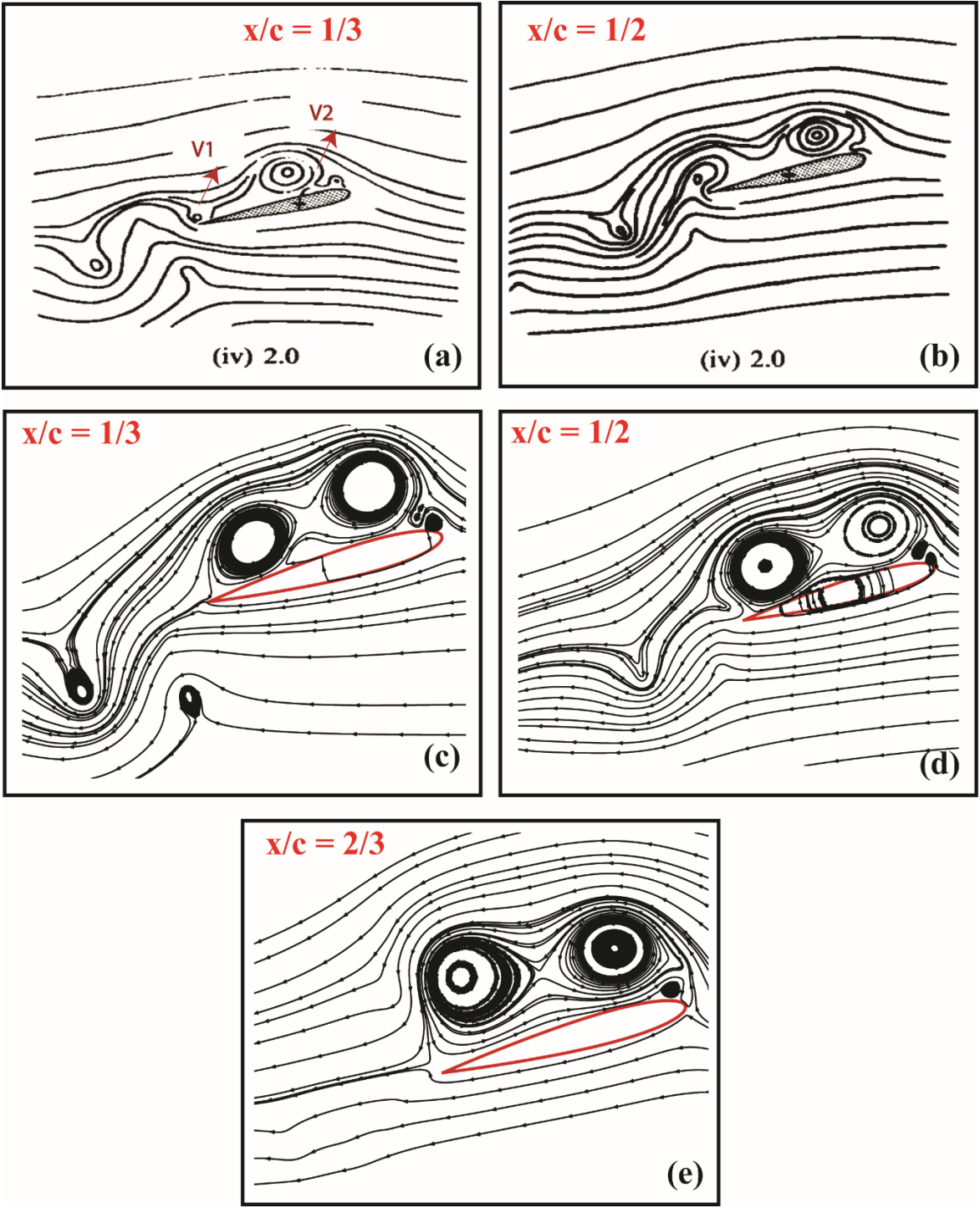

**Figure 16:** Streamline pattern for time instance t=2.0T. (a,b) experimental observation Ohmi, et al. (1991). (c,d,e) Results corresponding to three different pitching location (x/c)

At t = 1.0T, the streamline plots from Fig.14a show the presence of multiple leading edge vortices for the case with pitch axis location at one-third of the chord. The wake region indicates the presence of two vortices of opposite orientation. The strength of the vortex shed from the lower surface is higher than that of the vortex shed from the upper surface. When the pitch axis location moves down to half chord (Fig.14b), only one big, leading-edge vortex is visible. In the wake only one vortex that is shed from the lower surface is visible. The one shed from the upper surface is disintegrated when it reaches the end of its down-stroke. The results from the present simulation (Fig.14c-d) matches well with the experimental observation. For the case whose pitch axis location is at two-thirds of the chord, the starting vortex disintegrates, and only a single leading edge vortex is visible.

At t = 1.5T (Fig. 15a), the leading edge vortex formed is being convected downstream along the surface and formation of new leading edge vortex. The second trailing edge vortex shed from the lower surface of the airfoil is observed in the wake. This corresponds to a case whose pitch axis is at one-third of the chord.

Fig.15b (pitch axis located at half chord) shows that here too, leading-edge vortex convects downstream, but no new leading-edge vortex is formed yet. The second shed trailing edge vortex is also observed, but its strength is lesser compared to the former case. Fig. 15c-d shows the results obtained from the present simulation. The vortical structures found in the experiments are captured well. Fig. 15e corresponding to the case, whose pitching axis is located at two-thirds of the chord, shows that only leading-edge vortex is present and hasn't convected as much as the other two cases.

The vortex (V1) (Fig.15a) which is convecting downstream during the downstroke, loses its strength as it reaches the trailing edge at t=2.0T (Fig. 16a). During this time interval, the leading-edge vortex (V2) too has grown in its size and strength. Formation of the new vortex is also observed at the leading edge. In the wake two shed vortices with opposite orientations are seen. Fig.16b also provides similar observation, but the vortex on the trailing edge surface is stronger than the earlier case as it has just begun interacting with the boundary layer. Figs.16c-d shows that the present simulation captures the above experimental observations reasonably well. Fig.16e also exhibits similar leading-edge vortex dynamics but shows no trailing edge vortex formation.

In summary, the dynamic stall phenomena of pitching airfoil are simulated using a sharp interface immersed-boundary approach. A detailed study analyzing the effect of pitching axis location on the vortex dynamics is carried out. It is observed that moving the pitching axis downwards primarily, effects the trailing edge vortex dynamics. Evolution of leading-edge vortex remains largely unaffected. The simulated results show good agreement with the experimental results of Ohmi et al. (1991)

## 4. Conclusion

Flow past both stationary and moving airfoil has been successfully simulated using a sharp interface immersed-boundary method. The issue of spurious oscillation and mass conservation are handled through a field extension strategy that extends the flow field into the ghost nodes beneath the immersed surface. Validation result of the inline oscillating cylinder is presented to show that the schemes used to produce results that match qualitatively and quantitatively with experimental results. The flow physics of pitching airfoil is investigated by studying its wake structure, vortex dynamic evolution, the time scale of the events. The observations show good agreement with the experimental results. In the future, this framework can be extended to study flow physics involving more realistic geometry and complicated motions.

## Acknowledgments

Simulations are carried out on the computers provided by the Indian Institute of Technology Kanpur (IITK) (www.iitk.ac.in/cc) and the manuscript preparation as well as data analysis has been carried out using the resources available at IITK. This support is gratefully acknowledged.